\documentclass[twocolumn,showpacs,preprintnumbers,amsmath,amssymb,floatfix]{revtex4}

\usepackage{graphicx}
\usepackage{dcolumn}
\usepackage{bm}

\newcommand{\beq}{\begin{equation}}
\newcommand{\eeq}{\end{equation}}
\newcommand{\bey}{\begin{eqnarray}}
\newcommand{\eey}{\end{eqnarray}}

 \newcommand{\kms}{\, {\rm km \, s}^{-1} }
\newcommand{\mss}{\, {\rm m \, s}^{-2} }

\begin{document}

\preprint{APS/123-QED}

\title{Unifying all mass discrepancies with one effective gravity law?}

\author{HongSheng Zhao} \affiliation{SUPA, School of Physics and Astronomy, University of St Andrews, KY16 9SS, Fife, UK \\ Vrije Universiteit, De Boelelaan 1081, 1081 HV Amsterdam, Netherlands}

\author{Benoit Famaey} 
\affiliation{Observatoire Astronomique,  Universit\'e de Strasbourg, CNRS UMR 7550, Strasbourg, France}

\date{\today}

\begin{abstract}

A remarkably tight relation is observed between the Newtonian gravity sourced by the baryons and the actual gravity in galaxies of all sizes. This can be interpreted as the effect of a single effective force law depending on acceleration. This is however not the case in larger systems with much deeper potential wells, such as galaxy clusters. Here we explore the possibility of an effective force law reproducing mass discrepancies in all extragalactic systems when depending on both acceleration and the depth of the potential well.  We exhibit, at least at a phenomenological level, one such possible construction in the classical gravitational potential theory. Interestingly, the framework, dubbed EMOND, is able to reproduce the observed mass discrepancies in both galaxies and galaxy clusters, and to produce multi-center systems with offsets between the peaks of gravity and the peaks of the baryonic distribution. 

\end{abstract}

\maketitle

\section{Introduction}

The dynamics of spiral galaxies exhibit a well-documented fine-tuned conspiracy between the surface density of baryonic matter and the total gravitational field $|\nabla \Phi|$ \cite{FM}. Whatever the explanation for it, this observational fine-tuned relation involves an acceleration scale $a_0 \sim cH_0/2\pi \sim 10^{-10} {\rm m} {\rm s}^{-2}$ such that the effects of the putative dark matter appear at $|\nabla \Phi|<a_0$ and disappear at  $|\nabla \Phi|>a_0$ in galaxies of all sizes.  This is encapsulated by the empirical formula of Milgrom \cite{Mil83}, at the basis of the Modified Newtonian Dynamics (MOND) paradigm. 

Nevertheless, in the central parts of galaxy clusters, where the observed acceleration $|\nabla \Phi| > a_0$, the above prescription underpredicts the mass discrepancy by a factor of a few. Putting aside the obvious possibility of this constituting a practical falsification of MOND, three other possibilities are that (i) there are dark baryons in the central parts of clusters, (ii) there is additional non-baryonic dark matter in the central parts of clusters, or (iii) Milgrom's formula is the limiting case of a more general empirical relation. For extensive discussions of (i) and (ii) see, for instance \cite{Milgrom08, Sanders99,Sanders03,Angusbullet,AFD,Natarajan}. Here we rather concentrate on the possibility of case (iii).

A possibility to explain the discrepancy in galaxy clusters is that the effective force law needs a new scale in addition to $a_0$. An obvious scale distinguishing galaxy clusters from galaxies is the deepness of the potential well $|\Phi|$. For instance, Bekenstein \cite{Bek} recently proposed in this vein to add a velocity scale to Milgrom's prescription, such that the acceleration scale $a_0$ is a function of $|\Phi|$. In this note, we explore this very general idea and digress on its observational consequences. We start by reviewing the MOND formalism in Sect.~2, then expose its generalization in Sect.~3. Plausible variations of $a_0$ as a function of $\Phi$ and their observational consequences are then explored in Sect.~4 and 5 respectively, and conclusions are drawn in Sect.~6.

\section{MOND}

The idea of MOND is to link the observed gravitational attraction, $\vec{g} = - \nabla \Phi$, to the Newtonian gravitational field calculated from the observed distribution of visible matter, $\vec{g_N} = - \nabla \phi_N$, by means of an interpolating function $\mu$:
\begin{equation}
\mu\left(\frac{|\vec{g}|}{a_{0}}\right)\vec{g} \approx \vec{g}_{N},
\label{eq:A}
\end{equation}
where
$\mu(x) \rightarrow x$ for $x \ll 1$ and
$\mu(x) \rightarrow 1$ for $x \gg 1$. However, this expression is not robust since it does not respect usual conservation laws. A consistent modification of gravity at the classical level should come from modifying the gravitational part of the Newtonian lagrangian density
${\cal L}_{\rm Newton}= - \rho \phi_N - |\nabla \phi_N|^2/(8 \pi G)$. Bekenstein \& Milgrom \cite{BM84} have developed a modified gravity framework where $|\nabla \phi_N|^2$ is replaced by $a_0^2 F(|\nabla \Phi|^2/a_0^2)$, so that the modified lagrangian density reads:
\begin{equation}
{\cal L}_{\rm MOND} \equiv -  \rho \Phi - \frac{a_0^2}{8 \pi G} F\left( x^2 \right),
\label{aqualaction}
\end{equation}
where $x = \frac{|\nabla \Phi|}{a_0}$, and $F$ can {\it a priori} be any dimensionless function. Varying the action with respect to $\Phi$ then leads to a non-linear generalization of the Newtonian Poisson equation:
\begin{equation}
\rho = \nabla . \left[ \frac{\mu \left( x \right)}{4 \pi G}  \nabla \Phi \right]
\label{BM}
\end{equation}
where $\mu(x) = F'(x^2)$. In order to recover the right limits for $\mu(x)$, one needs to choose $F(y) \rightarrow y \; {\rm for} \; y \gg 1 \; {\rm and} \; F(y) \rightarrow \frac{2}{3}y^{3/2} \; {\rm for} \; y \ll 1.$

In the most popular covariant version of the theory \cite{teves,ZF,faprd} the weak-field potential is $\Phi = \phi_N+\phi$ where $\phi$ is a scalar field whose lagrangian density is proportional to $a_0^2F\left(a_0^{-2}h^{\alpha\beta}\nabla_\alpha \phi \nabla_\beta\phi\right)$, where $h^{\alpha\beta}$ is a combination of the metric and of a time-like unit vector. This scalar field lagrangian density is thus fully similar to Eq.~\ref{aqualaction}, and any modification of the classical MOND lagrangian could thus immediately be translated into a modification of the relativistic version of the theory.

\section{A more general force law?}

Eqs.~\ref{aqualaction} and~\ref{BM} suggest that the gravitational constant $G$ is effectively a running function of the gradient modulus $|\nabla \Phi|$ only. In other words, the lagrangian of the associated scalar field $\phi$ has no potential and depends only on its kinetic term. Any direct dependency on the field $\Phi$ would however be natural, since it is more general and, as we shall see, empirically desirable.

As noted by Bekenstein \cite{Bek}, an obvious distinction between galaxies and galaxy clusters is the deepness of the potential well. So one could imagine that the effective transition-acceleration $a_0$ is not a constant but is a monotonically increasing function of the potential depth, e.g., $\exp{|{\Phi/(1000\kms)^2}|} \times 10^{-10}\mss$, which boosts the MOND effect in galaxy clusters by a factor of a few\cite{Bek}.  However, 
a side effect of Bekenstein's exponentially-varying function is that it predicts a value of $a_0$ of the order of $10^{10000} \mss$ once the potential reaches the order of $c^2$, i.e.,  a neutron star or a stellar black hole would exhibit an undesirable deep-MOND behavior (see Fig.~\ref{figA0}).  
Its effect on large scale structures can be severe, too.  
It thus seems necessary to study a wider set of models for the effective variation of $a_0$, and select the best one from a phenomenological point of view.  

To extend the MOND lagrangian in this vein a very general way is to write the following lagrangian:
\begin{equation}
{\cal L}_{\rm EMOND} \equiv - \rho \Phi - \frac{\Lambda}{8 \pi G} F(x^2), ~ \Lambda \equiv A_0(\Phi)^2,
\label{LEMOND}
\end{equation}
where $x^2=y=\frac{|\nabla \Phi|^2}{\Lambda}$, and 
$A_0(\Phi)$ plays the role of $a_0$ in the usual MOND.  
This leads to the following modified Poisson equation, generalizing the MOND equation (Eq.~\ref{BM}):
\begin{equation}
\rho = \underbrace{\nabla . \left[ \frac{\mu \left(x\right)}{4 \pi G} \nabla \Phi \right]}_{T_1} - \underbrace{\left|\frac{\partial_{\Phi} \Lambda}{8\pi G}\right| F_1 \left(x^2\right)  }_{T_2}
\label{emond}
\end{equation}
where $\mu(x)=F'(x^2)$, and $F_1(y)  \equiv y F'(y) -F(y) = \int_0^\mu y d\mu$, which is typically a positive quantity of order unity.

Interestingly, with this ``extended MOND'' (EMOND) formalism, Gauss' theorem (or Newton's second theorem) would no longer be valid in spherical symmetry. This means that masses outside a sphere can affect the local gravitational field strength, and the empirical Eq.\ref{eq:A} no longer holds rigorously even in spheres, due to the additional term $T_2$, although this term is often small as we shall see in the next sections. So we still have approximately the EMOND counterpart of Milgrom's empirical law: 
\begin{equation}
\mu\left(\left|\frac{\nabla \Phi}{A_0(\Phi)}\right|\right)\nabla \Phi\approx \nabla \Phi_N.
\label{eq:B}
\end{equation}
Masses outside a spherical shell can also affect the local gravitational field strength by changing the boundary value of $\Phi$ or the zero-point of the potential $\Phi_{\infty}$, hence changing $A_0$ at the shell.  The appearance of such an absolute zero-point is a curable artifact of our introduction of a non-covariant $A_0(\Phi)^2$ instead of a covariant $A_0(\phi)^2$. Our lagrangian in Eq.~\ref{LEMOND} can be easily redesigned for a covariant scalar field, e.g., by replacing $-A_0(\Phi)^2 F(|\nabla \Phi|^2/A_0(\Phi)^2) \rightarrow -A_0^2(\phi)F\left(A_0(\phi)^{-2}h^{\alpha\beta}\nabla_\alpha \phi \nabla_\beta\phi\right)$.  The boundary values of $\phi$ and $A_0(\phi)^2$ are then well-defined by the cosmological evolution of $\phi_\infty$. The zero point of the Newtonian potential $\phi_N$ can raise or lower $\Phi$, but this has no effect on the scale $A_0^2(\phi)$ or the acceleration or the light deflection power $\nabla \Phi = \nabla \phi + \nabla \phi_N$. Note that we hereafter assume that the covariant version deflects light in a similar fashion as General Relativity, by invoking the same vector field as in TeVeS \cite{teves} in addition to the scalar field $\phi$ \footnote{The simplest scalar lagrangian could be $-(h^{\alpha\beta}\nabla_\alpha \phi \nabla_\beta\phi)^{3/2} A_0(\phi)^{-1}$, where $A_0 \propto |\phi|^{1/6}$ can increase from $a_0$ in galaxies to $2a_0$ in clusters and to $10a_0$ in black holes for $(300\kms)^2 < |\phi| < c^2$.}

\section{A realistic effective variation of the acceleration scale}

\begin{figure}
\centerline{\includegraphics[angle=0,width=9cm]{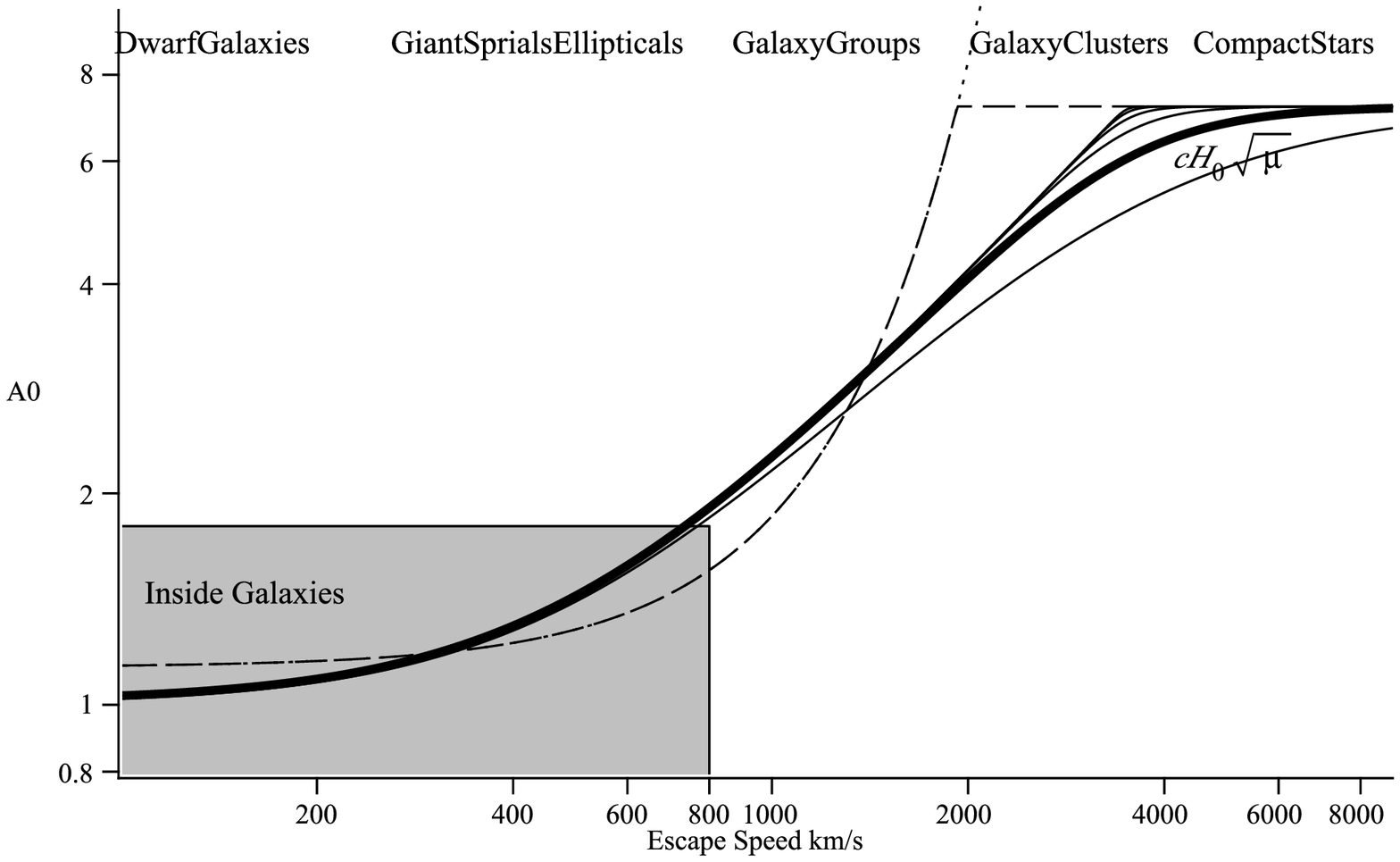}}
\centerline{\includegraphics[angle=0,width=9cm]{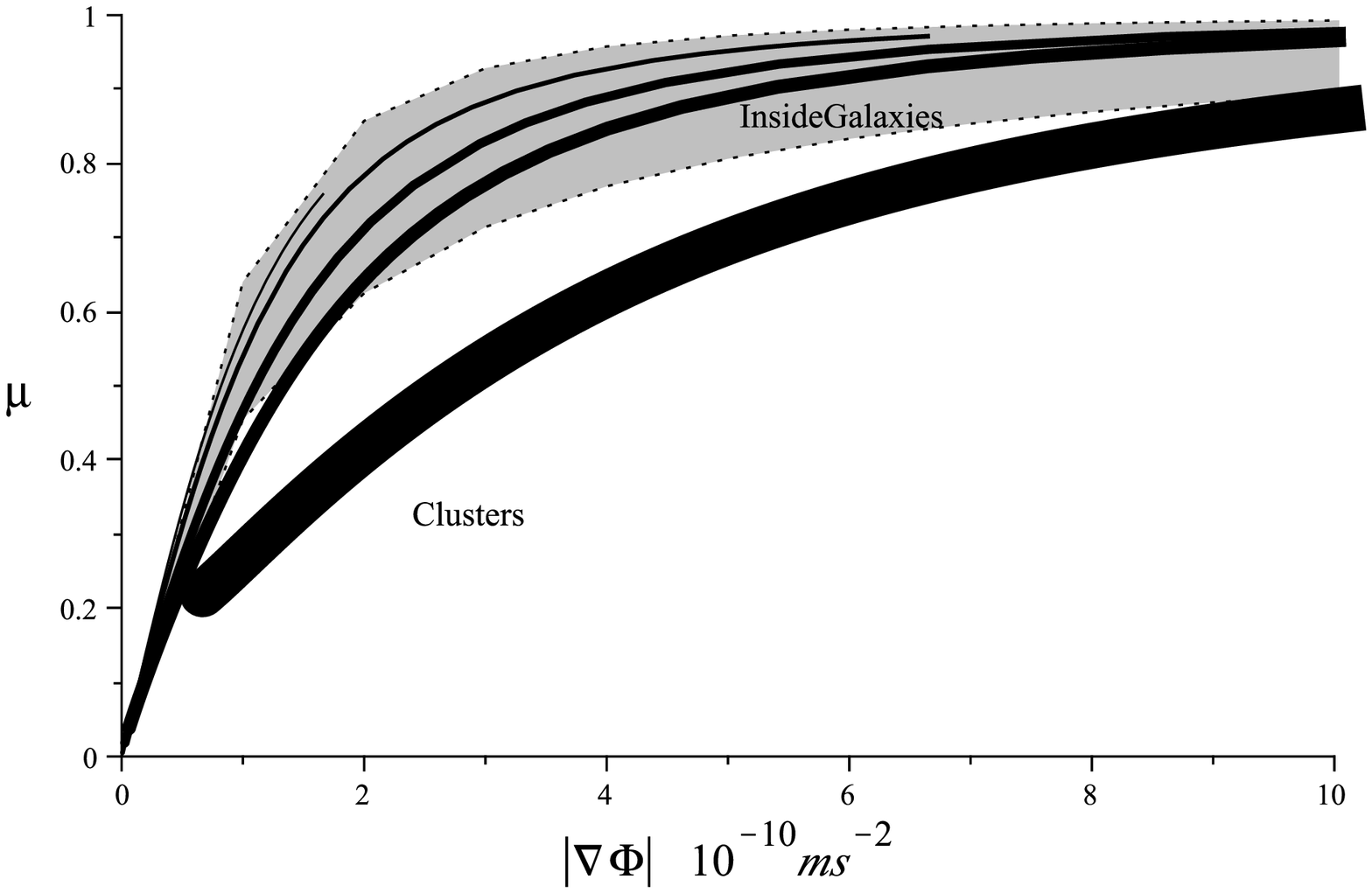}}
\caption{Top: 
shows the $A_0(\Phi)$ in units of $10^{-10}\mss$ 
for systems of increasing escape speed $V_{\mathrm esc} = \sqrt{2\Phi- 2\Phi_{\infty} }$ 
for several possible functions: $A_0(\Phi) = cH_0 \sqrt{ \mu(|\Phi|/(2500)^2)}$ (Eq.~\ref{mulike}) for $\mu(x)=x (1+x^n)^{-1/n}$ 
with $n=2$ (thick solid), $n=1, 4, \ldots, \infty$ (thin solid curves). Also shown are the truncated curve 
$A_0(\Phi) = {\rm exp}([|\Phi/(1000)^2|,~2 ]_{\rm min}) \times 10^{-10}\mss$ [dashed] and the corresponding exponential untruncated curve [dotted].  
Bottom: The value of $\mu(|\nabla \Phi|/A_0(\Phi))$ vs. $|\nabla \Phi|$ using Eq.~\ref{mulike} for systems with circular velocities of $1000$ km/s (galaxy clusters, thickest curve) as well as $300$, $200$, $100$ and $50$ km/s (thin curves); the latter four are in the grey zone between the traditional forms (dotted curves) $\mu(x)=x/(1+x)$  and $\mu(x)=x/(1+x^2)^{1/2}$, acceptable in all galaxies.}\label{figA0}
\end{figure}

Keeping the same philosophy as the {\it a priori} unknown $\mu$-function of MOND in mind, we examine plausible potential dependencies of $A_0(\Phi)$ from a phenomenological point of view. The only phenomenological requirements would be that (i) $A_0$ does not have a too large value in the strong field or cosmology regime, such that General Relativity is effectively recovered there, (ii) $A_0$ takes roughly the usual value of $a_0$ in galaxies and does not vary too much from galaxy to galaxy or inside a given galaxy, (iii)  $A_0$  is boosted in galaxy clusters by a reasonable amount compared to the usual value of $a_0$ so to make a real observational difference with MOND. 

Note also that isolated and non-isolated systems should be coherently transitioned from each other. This means that the potential $\Phi$ (or scalar field $\phi$) in $A_0$ should really be the total potential due to all the material. This reminds of the external field effect (EFE) of MOND \citep{FM}, but is more contriving since the external potentials can be non-neglibible even when the external force is negligible. So, the fact that one can neglect the weak EFE acting on some objects in MOND does not mean that one can ignore the external potential effect (EPE) of EMOND. The cosmological value of the scalar field in the covariant version would thus set a cut-off to $A_0(\phi)^2$.  A crude estimate of the EPE is to consider that most galaxies and dwarfs are actually within some $\sim 15$~Mpc of a galaxy cluster (e.g., Virgo is a rich cluster $\sim 15$~Mpc from us). Such clusters will typically contribute some $(300\kms)^2$ to the potential at the boundary of galaxies. So we can consider that the total $|\Phi|$ is typically always above $|\Phi_\infty| \sim 10^{-6} c^2$, which will set $A_0$ close to a lower cut-off with little variation for small internal potentials of small objects.  Similarly, $A_0$ is close to an upper cutoff for galaxies residing inside the deep potential of a rich cluster, consistent with the deep-MOND like behavior of these cluster galaxies \cite{Natarajan}.

While there are many plausible functional forms  of $A_0(\Phi)^2$ (see Fig.~\ref{figA0}), a simple transition can be achieved by a $\mu$-like dependency on the potential without introducing any extra function: 
\begin{equation}
\Lambda \equiv A_0(\Phi)^2 = (cH_0)^2 \mu(p) \, , \, p = |\Phi c^{-2}| \beta^{-2} \label{mulike}
\end{equation}
with $\beta \sim 8 \times 10^{-3}$. Here $\mu$ is the same function as whichever $\mu$-function of MOND, except that the argument is the potential instead of the acceleration.  To be specific we shall use $\mu(x)= x/(1+x^2)^{1/2}$ hereafter unless otherwise stated.  This empirical interpolating function simultaneously determines $A_0$ through Eq.~\ref{mulike}.  The covariant version could take 
$A_0(\phi)^2 \sim (cH_0)^2 \mu\left( |\phi/ (\beta c)^2| \right)$.

With this choice, all the above requirements on a possible realistic dependency of $A_0$ on the potential are met. (i) $A_0^2$ varies very little for large scale structure, and is almost equal to the observed amplitude of the cosmological constant. The value of $A_0$ is limited to $cH_0 \sim 8 \times 10^{-10}\mss$ even for black holes. (ii) Due to the EPE described above, the value of $A_0$ does not go to zero but levels off at $(cH_0) \sqrt{|\Phi_\infty|/ (\beta c)^2} \sim 10^{-10}\mss$, 
i.e., the classical value of $a_0$ (see Fig.~\ref{figA0}) in small galaxies and in the solar system \cite{galia}. Additionally, there will be only very mild ($<50\%$) variations of $A_0$ inside galaxies, as well as between galaxies, small variations which might be desirable to explain some anomalies in strong lensing galaxies \cite{ZBTH,CZ,Chiu,Shan,FZarc}. 
In any case this scatter is acceptable as the MOND formula works for all galaxies as long as $a_0$ is within the range between 0.9 and $1.7 \times 10^{-10}\mss$ \cite{faprd,things}.  As illustrated on Fig.~1 (bottom panel), the effective $\mu$-function stays within the range allowed by data on galaxy scales.  (iii) Finally, the much smaller $\mu$ on scales of galaxy clusters (bottom panel of Fig.~\ref{figA0}) means a boost of MOND effect in these (see also next section).

As for the $T_{2}$ term in Eq.~\ref{emond} further examinations show that 
$
T_{2}= \frac{H_0^2}{4 \pi G \beta^2} F_2, ~F_2= p F''(p^2) \left(x^2 F'(x^2) -F(x^2)\right). 
$
In the deep-MOND regime we found $F_2<0.1$ for typical forms of $\mu(p)$ and $\mu(x)$.  
So this extra density $T_{2}$ has a maximum of about $0.1 \beta^{-2} \sim 1000$ times above the cosmic critical density. Such a low over-density is not very significant even at large radii in galaxies, and is mild in galaxy clusters.  

\section{Boost of gravity and offsets in galaxy clusters}

In order to illustrate how such an effective gravity law would affect the dynamics of galaxy clusters, we concentrate hereafter on the above $\mu$-like dependency (Eq.~\ref{mulike}). There are two basic questions to be answered: (i) how much dark matter would be inferred from a Newton-Einstein framework for a given distribution of baryons? (ii) Can this effective ``phantom" dark matter be generically offset from the baryons as observed in weak lensing maps of the Large Scale Structure and in objects such as the bullet cluster \cite{Angusbullet}?

\begin{figure}
\centerline{\includegraphics[angle=0,width=7.5cm]{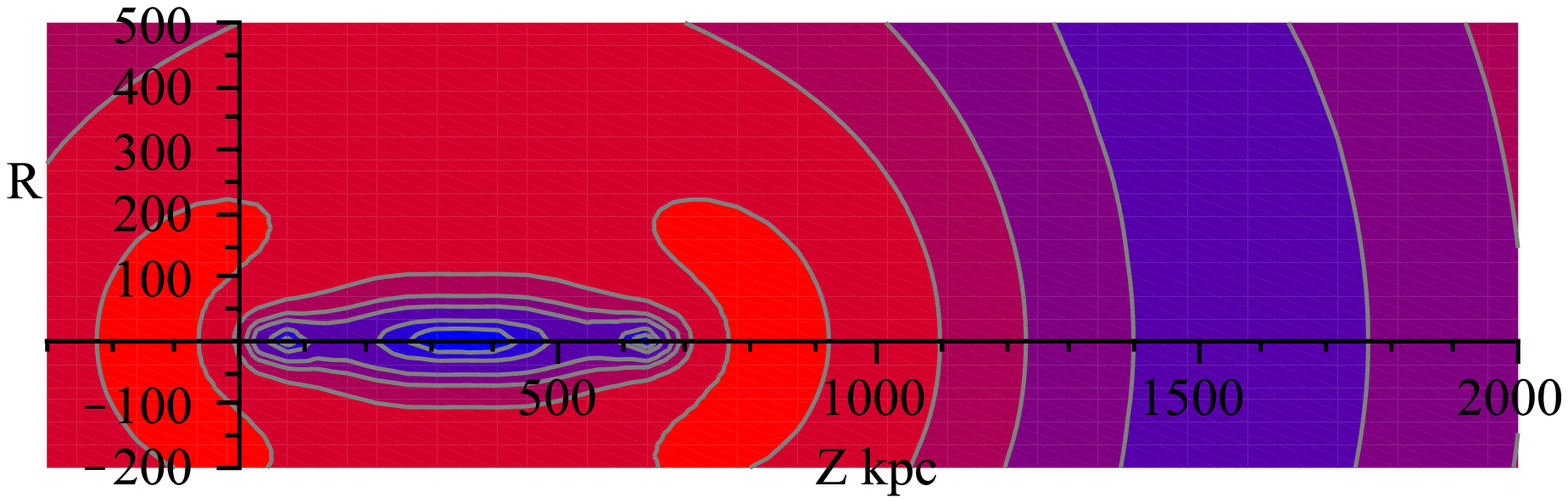}}
\centerline{\includegraphics[angle=0,width=7cm]{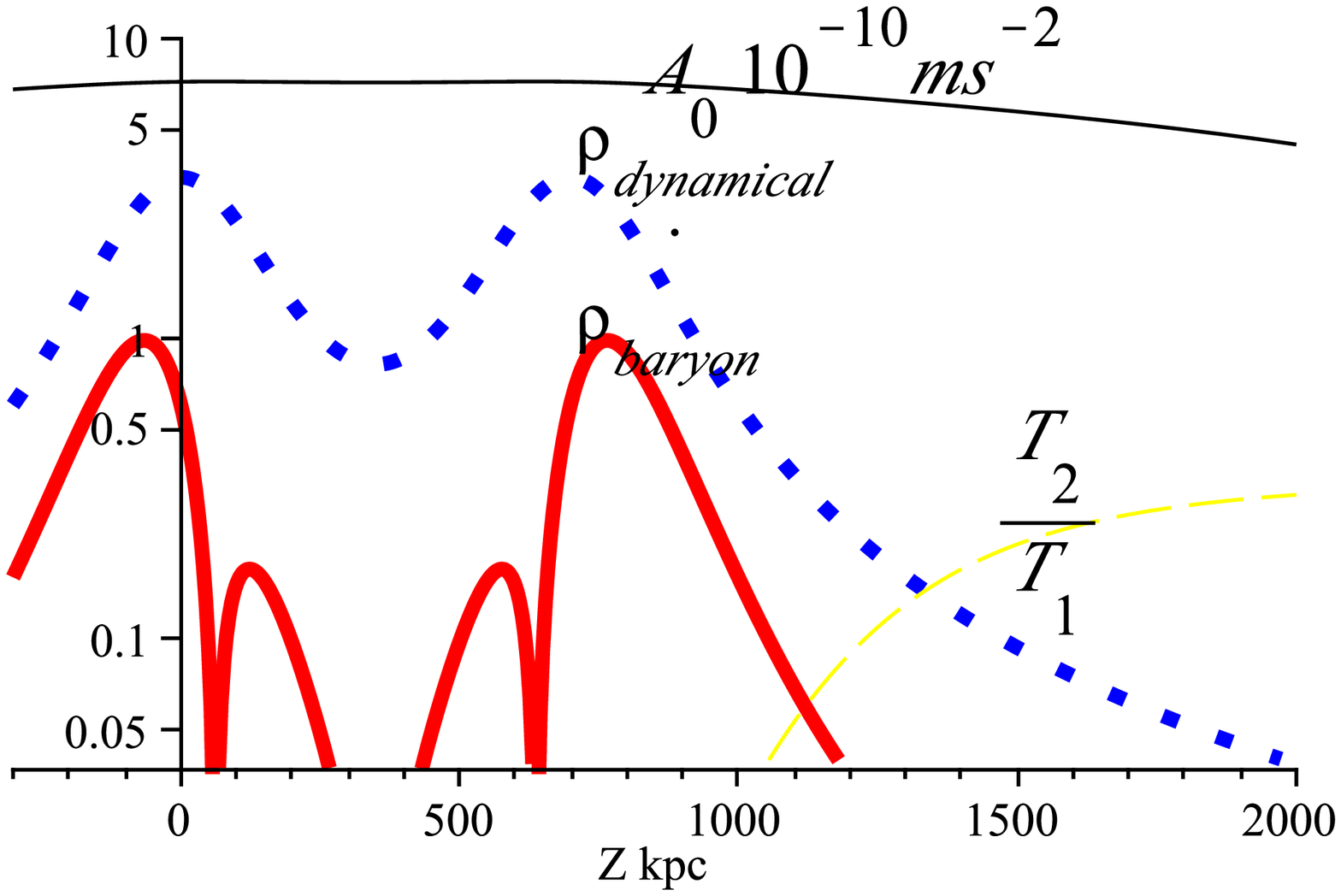}}
\caption{Top: Displays for the multi-center galaxy cluster the density ratio of EMOND baryonic mass over Newtonian dynamical mass 
$\rho_{\mathrm baryon}/\rho_{\mathrm dynamical} \sim \mu \sim |\nabla \Phi|/A_0 \sim 1/3$ (red, where the gravity peaks), 
$\sim 1/30$ (blue, where gravity is nearly zero). 
Bottom: Shows along the long axis $Z$,  the acceleration scale $A_0(\Phi)$ (thin solid), 
$\rho_{\mathrm dynamical} \equiv \nabla^2 \Phi/4\pi G$ (dotted), and $\rho_{\mathrm baryon} \equiv T_1-T_{2}$ (thick solid, see Eq.~5) in $10^6 M_\odot {\rm kpc}^{-3}$. 
The ratio $T_{2}/T_1$  (long-dashed) is shown to be small. }
\label{offset}
\end{figure}

In order to answer both questions, we consider a double peaked oblate potential:
$
\Phi(R,z) = \Phi_{\infty} + \sum_{i=1}^{i=2}
  1500^2 \ln \sqrt{r_i^2+200^2}/2500 ,
$
where $r_1 = \sqrt{z^2+R^2}$ and $r_2=\sqrt{(z-700{\mathrm kpc})^2+R^2}$.
We then determine the dynamical density distribution $\nabla^2 \Phi/4\pi G$ in Newtonian interpretation, and we determine the corresponding distribution of baryons in EMOND with Eq.~5.  The result is plotted on Fig.~\ref{offset}: the distribution of baryons is typically about 7 times less massive than the dynamical mass inferred from the same potential in Newtonian gravity, and interestingly, this distribution of baryons is 4-peaked, with the main baryonic peaks clearly offset from the corresponding peaks in the potential, or Newtonian dynamical mass peaks (see Fig.~\ref{offset}). It is thus generally possible to create large ``phantom" dark matter peaks generically offset from the baryon distribution within the present gravitational framework. This could be relevant to the modelling of the dark core in the trainwreck cluster \cite{trainwreck} or to the bullet cluster \cite{Angusbullet}, provided that the covariant version deflects light in a similar fashion as General Relativity, by e.g. invoking the same vector field as in TeVeS. We also note that the main baryonic peak seems to be generically in the more outer parts of the system compared to the main gravity peak, which is somewhat more similar to the situation of the train wreck cluster than to the bullet cluster. 

\section{Conclusions}
The effective gravity law explored in this note is able to boost gravity in galaxy clusters, eliminating the need for additional hot dark matter in MOND, a problem that has been known for more than 10 years \cite{Sanders99,Bek}. Generally speaking, the new framework actually needs two interpolating functions, and two fundamental constants defining their transition domains. The framework is in principle very flexible, and the details of the interpolating functions can be narrowed (or excluded) by observational constraints. It will for instance be important to test which forms can reconcile a large discrepancy in the clusters themselves with a not-too-large discrepancy in the cluster galaxies. Here, we recycled the standard $\mu$ function to tailor the $\Lambda=A_0^2$ function (Eq.~\ref{mulike}): the result is a transition between two plateaus,  $A_0^2 \sim (cH_0)^2$ for large scale structures and $A_0^2 \sim a_0^2$ for small galaxies.  To make the scheme (Eq.~\ref{LEMOND}) covariant and valid for lensing is straightforward (see end of Section 3). Future detailed dynamical and lensing studies of galaxies and galaxy clusters should allow one to test whether this ``extended MOND" (EMOND) scheme can indeed provide an effective gravity law reproducing mass discrepancies at all scales.

\end{document}